\documentclass[sigconf,screen,nonacm]{acmart}

\usepackage{amsmath, amsfonts}
\usepackage{graphicx}
\usepackage{textcomp}
\usepackage{xcolor}
\usepackage{tabularx}
\usepackage{microtype}
\usepackage{relsize}
\usepackage{todonotes}
\usepackage{subcaption}
\usepackage[normalem]{ulem}
\usepackage{booktabs}
\usepackage{hyperref}
\hypersetup{
    colorlinks=true,
    linkcolor=black,
    filecolor=black,
    urlcolor=black,
}
\usepackage{cleveref}
\crefname{section}{Sec.}{sections}
\Crefname{section}{Section}{Sections}
\crefname{figure}{Fig.}{figures}
\Crefname{figure}{Figure}{Figures}
\Crefname{table}{Table}{Tables}
\crefname{table}{Table}{Tables}

\usepackage{threeparttable}
\usepackage{pgf-umlsd} 

\usepackage{tikz}
\usepackage{pgfplots}

\usepackage{balance}

\usepackage{xspace}

\newcommand\parhead[1]{\vspace{.26mm}\noindent\textbf{{#1.}}}

\newcommand{\iI}[1]{#1\xspace}
\newcommand{\intervieweeI}[1]{#1\xspace}
\newcommand{\intervieweeII}[1]{#1\xspace}

\newcommand{\meier}{\intervieweeI{I\textsubscript{1}}}
\newcommand{\zang}{\intervieweeI{I\textsubscript{2}}}
\newcommand{\goerke}{\intervieweeI{I\textsubscript{3}}}
\newcommand{\klaas}{\intervieweeI{I\textsubscript{4}}}
\newcommand{\rasbach}{\intervieweeI{I\textsubscript{5}}}
\newcommand{\voss}{\intervieweeI{I\textsubscript{6}}}
\newcommand{\torkuhl}{\intervieweeI{I\textsubscript{7}}}
\newcommand{\zimmer}{\intervieweeI{I\textsubscript{8}}}
\newcommand{\runschke}{\intervieweeI{I\textsubscript{9}}}
\newcommand{\wilde}{\intervieweeII{I\textsubscript{10}}}
\newcommand{\mueller}{\intervieweeII{I\textsubscript{11}}}
\newcommand{\per}{\intervieweeII{I\textsubscript{12}}}
\newcommand{\meyer}{\intervieweeII{I\textsubscript{13}}}
\newcommand{\simon}{\intervieweeII{I\textsubscript{14}}}
\newcommand{\heppekausen}{\intervieweeII{I\textsubscript{15}}}
\newcommand{\kumbarji}{\intervieweeII{I\textsubscript{16}}}

\newcommand{\iempty}[1]{\textit{``#1''}}  
\newcommand{\imeier}[1]{\textit{``#1''}\,(\meier)}  
\newcommand{\izang}[1]{\textit{``#1''}\,(\zang)}  
\newcommand{\igoerke}[1]{\textit{``#1''}\,(\goerke)}  
\newcommand{\iklaas}[1]{\textit{``#1''}\,(\klaas)}  
\newcommand{\irasbach}[1]{\textit{``#1''}\,(\rasbach)}  
\newcommand{\ivoss}[1]{\textit{``#1''}\,(\voss)}  
\newcommand{\itorkuhl}[1]{\textit{``#1''}\,(\torkuhl)}  
\newcommand{\izimmer}[1]{\textit{``#1''}\,(\zimmer)}  
\newcommand{\irunschke}[1]{\textit{``#1''}\,(\runschke)}  
\newcommand{\iwilde}[1]{\textit{``#1''}\,(\wilde)}  
\newcommand{\imueller}[1]{\textit{``#1''}\,(\mueller)}  
\newcommand{\iper}[1]{\textit{``#1''}\,(\per)}  
\newcommand{\imeyer}[1]{\textit{``#1''}\,(\meyer)}  
\newcommand{\isimon}[1]{\textit{``#1''}\,(\simon)}  
\newcommand{\iheppekausen}[1]{\textit{``#1''}\,(\heppekausen)}  
\newcommand{\ikumbarji}[1]{\textit{``#1''}\,(\kumbarji)}  

\newcommand{\quotebox}[3]{\vspace{.5em}\noindent\begin{tikzpicture}
		               \node[align=center,draw,thin,minimum width=\columnwidth,inner sep=2.2mm] (titlebox)%
		               {\parbox{0.95\columnwidth}{\hspace{-4pt}\textit{#2\vspace{0.1cm}}}};
		               \node[label=left:{\colorbox{white}{\small #1}}] (W) at (titlebox.south east) {};%
		       \end{tikzpicture}\vspace{-12pt}}

\setcopyright{none}
\setcctype[4.0]{by}

\copyrightyear{2026}
\acmYear{2026}
\acmDOI{XXXXXXX.XXXXXXX}
\acmBooktitle{Companion Proceedings of the 34th ACM Symposium on the Foundations of Software Engineering (FSE '26), June 5--9, 2026, Montreal, Canada}
\acmConference{34th ACM Symposium on the Foundations of Software Engineering}{June 5--9, 2026}{Montreal, Canada}
\acmISBN{979-X-XXXX-XXXX-X/26/03}

\begin{document}

\title[Developer Perspectives on REST API Usability]{Developer Perspectives on REST API Usability:\\ A Study of REST API Guidelines}

\author{Sven Peldszus}
\orcid{0000-0002-2604-0487}
\affiliation{%
	\institution{IT University of Copenhagen}
	\country{Denmark}
}
\affiliation{%
\institution{Ruhr University Bochum}
\country{Germany}
}

\author{Jan Rutenkolk}
\author{Marcel Heide}
\author{Jan Sollmann}
\orcid{0009-0007-3973-4641}
\affiliation{%
	\institution{Ruhr University Bochum}
	\country{Germany}
}

\author{Benjamin Klatt}
\author{Frank Köhne}
\affiliation{%
	\institution{viadee AG}
	\city{Cologne}
	\country{Germany}
}

\author{Thorsten Berger}
\orcid{0000-0002-3870-5167}
\affiliation{%
	\institution{Ruhr University Bochum}
	\country{Germany}
}
\affiliation{%
	\institution{Chalmers$|$University of Gothenburg}
	\country{Sweden}
}

\begin{abstract}
	REST is today's most widely used architectural style for providing web-based services.
	In the age of service-orientation---a.k.a. Software as a Service (SaaS)---APIs have become core business assets and can easily expose hundreds of operations.
	While well-designed APIs contribute to the commercial success of a service, poorly designed APIs can threaten entire organizations.
	Recognizing their relevance and value, many guidelines have been proposed for designing usable APIs, similar to design patterns and coding standards.
	For example, Zalando and Microsoft provide popular REST API guidelines. However, they are often considered as too large and inapplicable, so many companies create and maintain their own guidelines, which is a challenge in itself.
	%
	In practice, however, developers still struggle to design effective REST APIs. 
	To improve the situation, we need to improve our empirical understanding of adopting, using, and creating REST API guidelines.
	
	\looseness=-1
	We present an interview study with 16 REST API experts from industry. We determine the notion of API usability, guideline effectiveness factors, challenges of adopting and designing guidelines, and best practices. 
	%
	We identified eight factors influencing REST API usability, among which the adherence to conventions is the most important one.
	While guidelines can in fact be an effective means to improve API usability, there is significant resistance from developers against strict guidelines. 
	Guideline size and how it fits with organizational needs are two important factors to consider.
	%
	REST guidelines also have to grow with the organization, while all stakeholders need to be involved in their development and maintenance.
	Automated linting provides an opportunity to not only embed compliance enforcement into processes, but also to justify guideline rules with educational explanations.
\end{abstract}



\maketitle

\renewcommand{\shortauthors}{Peldszus et al.}

\section{Introduction}\label{sec:intro}
\noindent\looseness=-1
Web service APIs have become critical business drivers for internal operations and the monetization of software services.
Internal APIs connect an organization's IT systems, and external APIs allow other organizations to access services.
According to hundreds of developers surveyed\,\cite{stateofapis}, web service APIs are intensively used by organizations of any size.
Large organizations with more than 10,000 employees often use over 250 APIs, while smaller organizations already use up to 50.

\looseness=-1
Today, Representational State Transfer (REST) is the dominant architectural style for delivering web services\,\cite{stateofapis}.
Already in 2017, REST was used in 18,400 APIs (82\,\% of all known APIs)\,\cite{programmable_web_2017}, growing to 24,471 by 2022\,\cite{programmable_web_2022}.
RESTful services are stateless and well separated from their clients via a unified interface, typically using HTTP endpoints.
Their APIs can be relatively large, with 62\,\% of REST APIs providing between 11 and 100 operations, and 18\,\% even more\,\cite{buelthoff_restful_2019}.

While RESTful services enable loose client-server coupling and independent service development, this flexibility introduces new challenges.
Most importantly, for a service to be actually used, its API must be easy to use, especially when competing with similar services of other organizations.
Usability often suffers from inconsistent naming or usage across API parts, sometimes to the extent that the service no longer adheres to REST principles\,\cite{kotstein_which_2021}.

In practice, organizations try to address API usability through developer guidelines that describe best practices for API design.
These are similar to design patterns\,\cite{Fraser2006} or coding standards, such as those by the SEI CERT\,\cite{sei-cert}.
While there are many well-documented REST API guidelines available, inconsistencies in REST APIs persist.
HTTP traffic analyzed in 2016 
revealed that \textit{``implementation and usage of REST APIs---as well as that of Web services more in general---is still far from being a stable and consolidated discipline.''}\,\cite{rodriguez_rest_2016}. 
According to preliminary discussions with industry experts, this has been a recurring theme over the past decade and has not improved since 2016. REST API usability remains a major pain point.
It is unclear what social or technical factors explain these observations, whether unclear definitions, disagreements, or random errors hinder the creation of high-quality REST APIs, or whether the observed inconsistency is an accepted byproduct of the development process.
To address this gap, three research questions must be answered.
\vspace{1.7pt}

\looseness=-1
\noindent\textbf{RQ1:} \textit{What is the developers' understanding of REST API usability?}
To determine the impact of guidelines, it is essential to capture developers' general knowledge and understanding of RESTful services, with a focus on API usability.
We investigated what developers consider a usable REST API and what factors influence usability.

\noindent\textbf{RQ2:} \textit{What obstructs the successful adoption of API guidelines?}
\noindent
While REST API guidelines are considered and important means to develop usable APIs, they do not seem to be having their intended effect. We study the reasons, specifically how the quality and organizational aspects influence the adoption of guidelines.

\noindent\textbf{RQ3:} \textit{What challenges creating and tailoring API guidelines?}
\noindent\looseness=-1
Since existing guideline catalogs are often large, generic, and difficult to adopt, organizations often need to create custom guidelines, or tailor existing ones to the organization or concrete project.

\looseness=-1
We present an interview study with 16 REST experts with substantial experience developing and maintaining REST APIs for different companies and domains. We transcribed and analyzed the interviews using open-coding to answer our research questions, determining the notion of API usability, challenges of adopting and designing guidelines, as well as best practices to develop usable RESTful services. We provide an online appendix\,\cite{appendix:online} with further details (e.g., interview guide).

On a final note, REST API guidelines are \textit{``a giant topic under which you can summarize anything you want''} according to one of our experts. This underscores the complexity and diversity of REST API guidelines. 
In this light, our results contribute to understanding the role and success factors of REST API guidelines. We hope they facilitate an informed discussion of challenges, potential mitigation strategies, and the appropriate balance between developer freedom and API consistency.

\section{Background and Motivation}
\noindent
We now briefly introduce REST and motivate our study by discussing API usability and the guidelines landscape.

\parhead{Representational State Transfer (REST)}
REST is an architectural style for web services\,\cite{fielding_dissertation_2000} that encompasses six properties.
(1) REST architectures realize the client-server model\,\cite{Sinha1992} to enforce separation of concerns,
(2) but optionally, servers can provide code on-demand\,\cite{Carzaniga1997,Carzaniga2007} for client-side execution to minimize traffic.
(3) RESTful services are stateless, avoiding to maintain sessions or store request-specific data on the server, but requiring each client request to include all necessary information.
(4) This allows caching of responses and improves scalability and performance.
(5) Clients cannot detect intermediaries, e.g., layers enforcing security, between themselves and the server.
(6) Services are provided via a uniform API that may not reflect internal server data structures, allowing clients and services to evolve independently.

\looseness=-1
REST API uniformity includes a Uniform Resource Identifier (URI)\,\cite{uri} for all accessible information, and a Uniform Resource Locator (URL)\,\cite{url} for each service.
URLs for follow-up requests are provided on demand in responses (cf. hypermedia as the engine of application state (HATEOAS)\,\cite{Alarcon2010}).
All responses use a consistent data format across the REST API, which may differ from internal structures but is detailed enough to support data manipulation.

How to realize REST architectures is not standardized\,\cite{kotstein_which_2021}.
In practice, services are mostly provided via HTTP endpoints, i.e., URLs targetable via HTTP requests, that return JSON data.
However, REST does not specify concrete technologies to be used, so developers could use XML instead of JSON for data representations.
APIs are either provided from provider to consumer, client developers in this case, or through public API directories\,\cite{github-publicapis}.

Several REST API frameworks are used to implement RESTful services\,\cite{Murugan2024}.
Among them, OpenAPI/Swagger\,\cite{swagger} is the most widely used framework for HTTP-based REST APIs and allows, among other things, generating server or client code from a REST API specification in OpenAPI format or generating an OpenAPI specification from annotated source code\,\cite{openapi}.
The OpenAPI format is a JSON-based, human-readable and machine-processable structure, thus facilitating understanding of services without requiring access to source code, documentation, or network traffic analysis.

\parhead{REST API Usability}
With the success of RESTful services, corresponding APIs have grown large\,\cite{DiLauro2022}, and maintaining multiple, long-lived APIs has become common.
This suggests similar effects on REST API quality, i.e., usability, similar to software aging in traditional software\,\cite{Parnas1994}, i.e., the steady decay of software quality with changes.
This can negatively impact the usability and security of a REST service.
Furthermore, since REST is no official standard\,\cite{kotstein_which_2021} like SOAP, with specific rules and syntax, this leads to an unfocused field of API usability and deviation from REST principles is common in practice.
Often, REST becomes a synonym for any Web API utilizing URIs and HTTP\,\cite{kotstein_which_2021}.
In particular, API consistency 
is an issue, since REST provides no standard for naming conventions, data formats, or similar concepts.
Thus, collaborative development results in inconsistencies that impact usability.

\begin{figure}
	\centering
	\includegraphics[width=\columnwidth]{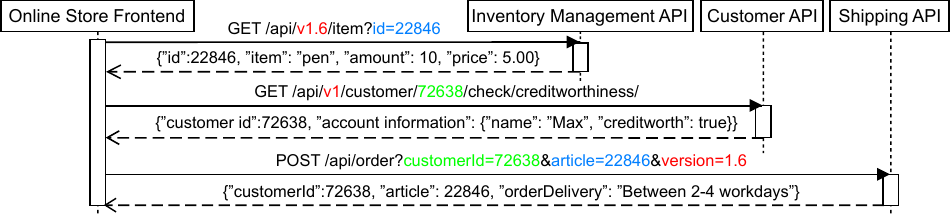}
%
%

	\vspace{-4pt}
	\begin{tiny}
		\begin{tabular}{@{}r@{}l@{}@{}r@{}l@{}}
			Inconsistencies:
			\textsuperscript{\color{red}\textbf{a}}\,versioning scheme;
			\textsuperscript{\color{green}\textbf{b}}\,parameter types;
			\textsuperscript{\color{blue}\textbf{c}}\,parameter or path id reference
		\end{tabular}
	\end{tiny}
	\vspace{-10pt}
	\caption{
		Illustration of inconsistently designed REST APIs
	}
	\label{fig:REST_SequenceDiagram}
	\vspace{-10pt}
\end{figure}

\looseness=-1
To illustrate the challenges of using an inconsistent REST API, \cref{fig:REST_SequenceDiagram} shows three consecutive API calls in a shop system, in which a front-end team must connect to APIs provided by different teams of the company.
The superscripts illustrate inconsistencies in the API designs:
{\color{red}a)} versioning schemes, i.e., major version only vs. minor version,
{\color{green}b)} inconsistent parameter types, i.e., path parameter vs. query parameter, and
{\color{blue}c)} inconsistent wording of semantically identical entities, i.e., id vs. name.
This challenges comprehension, maintenance, and evolution of APIs.




\parhead{REST API Guidelines}
\noindent\looseness=-1
To mitigate API quality problems, companies started creating API design guidelines to foster consistent API landscapes.
Guidelines provide examples and best practices for developers, and are often the basis for creating custom guidelines\,\cite{murphy_preliminary_2017}.
Such guidelines are typically a set of rules, each regarding different aspects in REST API development, i.e., one rule in Zalando's guideline in the category \textit{``REST Basics - Security''} is that REST APIs \textit{``MUST secure endpoints''}\,\cite{zalando}.
However, guidelines can vary widely because they are based on the experience of developers and the institutions in which they are created and
there are many guidelines---a GitHub search for ``REST Guideline'' in September 2025 returned 102 results.
More systematically, Wilde collected 27 guidelines and two compilations of guidelines \cite{dret:github}, including the API Stylebook\,\cite{apistylebook}.
Murphy et al. analyzed 32 public REST API guidelines and defined 27 aspect categories, e.g., naming conventions for URIs\,\cite{murphy_preliminary_2017}.
%

However, many REST APIs and their associated guidelines exhibit inconsistencies, concerning varying and sometimes conflicting approaches to key aspects such as documentation, error handling, and overall design standards\,\cite{murphy_preliminary_2017}.
Such inconsistencies may partly arise from linguistic antipatterns and poor documentation practices\,\cite{palma_semantic_2017}.
\looseness=-1
Implementing advanced REST concepts such as HATEOAS is not a priority for many developers\,\cite{kotstein_which_2021}.
Before, developers were more familiar with SOAP, which provides more structured guidance compared to the more flexible and loosely interpreted REST guidelines, resulting in low adoption of best practices and ideas of REST API development as captured in REST guidelines\,\cite{rodriguez_rest_2016}.
However, the current state of practice is unclear.
To enhance the standardization and usability of REST API guidelines, one suggestion is to develop more machine-readable resources\,\cite{wilde_surfing_2018}.

Overall, API guidelines are widely used, but it is questionable whether they achieve their goals.
It is unclear how developers perceive and use API guidelines, which needs to be understood to identify improvements and best practices.

\section{Related Work}
\noindent
Several studies address challenges in REST API design.
Bogner et al. experimentally confirm that adhering to RESTful API design rules significantly improves the understandability of APIs, and that violations clearly reduce comprehension regardless of user experience\,\cite{Bogner2023}.
Yamamoto et al. emphasize the need for automation to resolve behavioral discrepancies between web-based and local REST APIs, supporting consistency\,\cite{yamamoto_quality_2018}.
Daughtry et al. focus on improving API usability through clear and accessible documentation using tools like Swagger\,\cite{daughtry_uses_2017}.
In this context, Coblenz et al. show that while OpenAPI and static analysis tools support syntactic validation of REST APIs, developers face challenges in automating usability, context-aware design, and real-time feedback, underscoring the need for combining automated checks with manual, human-centered processes\,\cite{Coblenz2023}.
The Richardson Maturity Model has become a widely referenced standard framework for assessing REST APIs, from basic HTTP usage (Level 0) to advanced HATEOAS (Level 3)
\,\cite{maturitymodel}. 
Robles et al. analyze REST API evolution, showing that breaking changes decrease in newer versions, enhancing client-side stability\,\cite{robles2023exploratory}.

Nguyen et al.\cite{Nguyen2017} highlight REST API security concerns, noting the absence of standardized specifications compared to SOAP.
Iacono et al. underscore this need for robust security frameworks, especially in message-oriented architectures\,\cite{LoIacono2019}.
Abdulghani et al. propose security guidelines for IoT, pointing out gaps in standardized frameworks for REST API security, especially in resource-constrained environments\,\cite{Abdulghani2019}.
 Collectively, these studies call for comprehensive REST API standards that balance complexity, usability, and security. 
Work attempting to mitigate these challenges includes a proposed set of REST API design guidelines based on existing Web standards\,\cite{wilde_surfing_2018}. 
Koci et al. propose metrics for web API usability, expressed through the predicted error rate, and trained a classifier model to predict the error rates of API endpoints\,\cite{Koci2020}.
Verborgh and Dumontier\cite{verborgh_web_2018} explore feature-based reuse in web APIs, enabling shared client/server code, documentation, and tools to promote cross-API compatibility.
Daugthry et al. describe eight ways developers use interactive explorers for web APIs, grounded in empirical analyses of the Google APIs Explorer\,\cite{daughtry_uses_2017}.

Petrillo et al. survey the literature to extract a catalog of 73 best practices for REST API design and study well-known REST APIs (e.g., Google Cloud Platform) 
to determine how they are offered and accessed\,\cite{petrillo_are_2016}.
Similar to our study, Zhang et al. interview 23 API designers from 6 companies and 11 open-source projects to understand their practices and needs, but they focus on  user feedback through bug reports and peer reviews\,\cite{Zhang2020}.
In addition, Piccioni et al. present an API usability study on 25 programmers, which combines interview questions with systematic observations of programmer behavior while solving token-based programming tasks. So, their scope is limited to API design\,\cite{Piccioni2013}.

While related works demonstrate the advantages of adhering to REST API guidelines and address their specifics, the factors necessary for successfully adopting them remain unexplored, which is one of the aspects our study addresses.
\section{Methodology}
\noindent\looseness=-1
Our study required an exploratory, qualitative methodology able to uncover relevant factors.
Hence, we employed a respondent strategy\,\cite{Storey2020}, specifically an interview study. We conducted semi-structured interviews\,\cite{adams_william} with knowledgeable experts.

\parhead{Participant Selection}
We sampled interviewees via an industrial partner (a consulting company whose experts work for many companies) and its network, following a purposive sampling strategy \cite{Patton2015}.
Sampling criteria included professional expertise with REST APIs, diverse backgrounds, and different relations with REST APIs.
We ensured a range of seniority, from younger software developers to experienced software architects, as well as specialists and experts involved in creating popular guidelines and enterprise architecture tooling.
We selected both providers and consumers of REST APIs to incorporate perspectives on creating and using REST API guidelines.
We ensured a diverse range of backgrounds, including domains such as finance, insurance, telecommunication, commerce, and healthcare. 
Participants were recruited continuously until we reached saturation.
This way, we recruited 16 knowledgeable experts, whose experience, current role, relation to REST APIs, and professional focus (e.g., as consultants for different companies) is shown in \cref{tab:Interviewees}.
This number is within the expected range of 16 to 24 interviews for saturation in interview studies\,\cite{Hennink2016}.

\parhead{Interview Design}
Following the concept of intensive interviewing\,\cite{Charmaz2006}, we created an interview guide with broad, open-ended questions to facilitate dynamic and detailed discussions.

\textit{1) Background \& Project Context:} First, asking general questions about the participant, i.e, the role in the job, experience with REST, and the project context.
\textit{2) REST API Usability:} Second, about the developers' perception of what makes up a well-usable REST API (RQ1).
\textit{3) Experiences with REST Guidelines:} Only thereafter, to not bias the developers in their perception of REST API usability, we asked them about their experiences with REST guidelines to complete the collection of background information.
\textit{4) Guideline adoption:} We continued with questions on their experiences in using guidelines and asking about challenges in everyday use (RQ2).
\textit{5) Creating guidelines:} Lastly, we asked about experiences in tailoring existing guidelines or creating new ones (RQ3).

These broad questions were designed to leave the interviewer room for detailed follow-up questions and allowing to steer the interview into directions that have not been uncovered in previous interviews\,\cite{Charmaz2006,Kvale2009}.
This design aligns with qualitative research principles, where not every participant needs to answer the same questions\,\cite{Miles2014}. 
Instead, the interviewer adapts follow-up questions based on the participant’s responses and emerging themes, allowing for flexibility and depth in exploration.
Our appendix\,\cite{appendix:online} provides the interview guide and the questions asked in each interview.
Due to privacy concerns and the confidential nature of the interviewees’ and companies’ information, transcripts cannot be released.

\parhead{Interview Execution}
Each interview was conducted in German and lasted an average of 30 minutes. Earlier interviews tended to be longer, while later ones tended to be shorter due to response saturation.
Before the main interviews, a trial interview was executed to refine the interview guide, test clarity and relevance of the questions, and ensure a smooth interview process\,\cite{Charmaz2006,Kvale2009}.
After each interview, its recording was transcribed and analyzed before the next interview to account for new factors.
Analysis followed a rigorous coding process using MAXQDA\,\cite{MAXQDA}.
First, the 1{st}, 2{nd}, and 3{rd} authors of this paper individually performed open coding to extract all relevant aspects\,\cite{Boyatzis1998}, with the 1{st} author validating all codes after coding. In the next step, these code proposals were then aggregated and refined by the 1{st} and 4{th} authors based on collaborative discussions between the two authors\,\cite{Corbin2014,Williams2019}. Through multiple iterations, the two authors identified key aspects, refined them, and validated them with supporting interview snippets, ensuring a thorough and  transparent analysis capturing nuanced insights.

\looseness=-1

\begin{table}[]
	\caption{Overview of the Interviewees}
	\label{tab:Interviewees}
	\centering
	\setlength{\tabcolsep}{2pt}
	\smaller
	\vspace{-10pt}
	\begin{tabular}{l c l c c}
		\toprule
		\textbf{ID} & \textbf{Exp.}\textsuperscript{1} & \textbf{Role(s)} & \textbf{Relation to REST} & \textbf{Focus}\textsuperscript{2}\\ \midrule
		\meier  & 8 & developer, architect & provider & consultant\\ 
		\zang & -\textsuperscript{3} & consultant & provider  & consultant\\
		\goerke & 2 & data scientist & provider  & consultant\\
		\klaas & 4 & developer, architect & provider  & consultant\\
		\rasbach & 3 & developer& consumer  & consultant\\
		\voss & 10 & architect & consumer \& provider  & consultant\\
		\torkuhl & 2 & developer& consumer \& provider  & consultant\\
		\zimmer & 1.5 & architect & consumer \& provider & consultant\\
		\runschke & 5 & developer, consultant & consumer  & consultant\\ 
		\wilde & 10 & consultant, guideline author & provider & consultant\\ 
		\mueller & 8 & architect, guideline author & provider & company\\ 
		\per & 20 & manager & provider & company \\ 
		\meyer & 12 & architect & provider & company\\ 
		\simon & 5 & developer & provider & company\\ 
		\heppekausen & -\textsuperscript{3} & developer, architect & provider & company\\
		\kumbarji & 0.5 & developer & provider & company \\
		\bottomrule
	\end{tabular}

	\vspace{1pt}\scriptsize
	\textsuperscript{1}experience with REST in years;
	\textsuperscript{2}consultant: works for multiple companies (broad domain experience) / company: works for one company (focused domain experience);
	\textsuperscript{3}experience exists, but was not quantified
	\vspace{-14pt}
\end{table}

\parhead{Ethnography and Context}
\looseness=-1
To contextualize this study, we captured the interviewees' ethnographies and their work with REST APIs based on the responses in interview parts 1) and 3).
We also used these dimensions to identify differences in the perception of senior and junior developers, consumers and providers of APIs, and consultants and developers from their company customers.

As shown in \cref{tab:Interviewees}, the interviewees are primarily developers and software architects.
The majority has 5--10 years of experience with REST APIs, five less than five years, and two more than 10 years, while two did not specify their experience with REST.
Most use or develop REST APIs.
In addition, four interviewees manage or sell API endpoints without developing them, two are migrating legacy systems to REST, and one is using REST in process automation, internal training, and providing support services and software for API management.
Two interviewees did not specify their context for using REST.
Half of the interviewees are familiar with popular REST API guidelines, i.e., of Zalando\,\cite{zalando} or Microsoft\,\cite{microsoft}.
One even mentioned having used about 15 different guidelines.
In addition, three interviewees only knew their internal guidelines, three knew general best practices, and two did not know any guidelines.
\section{Usability of REST APIs (RQ1)}
\noindent\looseness=-1
To better understand the usability of REST APIs, we first surveyed the interviewees for their perspective on what makes a usable API, and then how to create usable REST APIs.

\subsection{Factors Influencing REST API Usability}
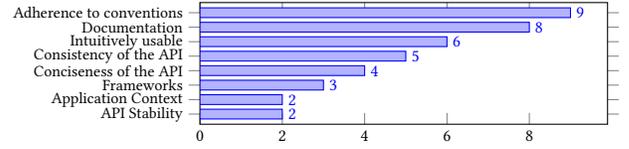
\begin{figure}
		\centering
		\scriptsize
		\begin{tikzpicture}
			\begin{axis}[
				xbar, xmin=0,
				symbolic y coords={API~Stability,Application~Context,Frameworks,Conciseness~of~the~API,Consistency~of~the~API,Intuitively~usable,Documentation,Adherence~to~conventions},
				ytick=data,
				nodes near coords, nodes near coords align={horizontal},
				bar width=0.13cm,
				height=3.2cm, 
				width=7cm,
				enlarge y limits=0.1, 
				]
				\addplot coordinates {(9,Adherence~to~conventions)(8,Documentation)(6,Intuitively~usable)(5,Consistency~of~the~API)(4,Conciseness~of~the~API)(3,Frameworks)(2,Application~Context)(2,API~Stability)};
			\end{axis}
		\end{tikzpicture}
		\vspace{-12pt}
		\caption{Factors influencing the usability of a REST API}
		\label{fig:factors}
		\vspace{-18pt}
	\end{figure}

\noindent\looseness=-1
To obtain a first picture, 
we asked the interviewees about their general understanding of REST API usability and what impacts it.
We identified 8 factors that influence REST API usability (see \cref{fig:factors}).

\parhead{Adherence to Conventions}
\looseness=-1
For the majority (9 out of 16 interviewees), it is essential that usable REST APIs build upon existing conventions (\iI{I\textsubscript{1}}, \iI{I\textsubscript{4-5}}, \iI{I\textsubscript{7-8}}, \iI{I\textsubscript{10}}, \iI{I\textsubscript{12-13}}, \iI{I\textsubscript{15}}), i.e., that \iklaas{conventions are used wherever possible}.
Particularly, hard to use APIs are often \izimmer{not a structured API but simply a web service that also uses JSON but is not defined according to the REST pattern}.
Four interviewees emphasized the resource-orientation of REST (\iI{I\textsubscript{1}}, \iI{I\textsubscript{12-13}}, \iI{I\textsubscript{15}}), stating that \imeier{R [in REST] stands for resource} and to expect
\iheppekausen{smaller services that work on a resource basis and do not deliver everything to me so that I get a huge conglomerate, but that I can work on a resource basis (...)}.
In particular, \iI{I\textsubscript{1}} states that HTTP and JSON are mandatory---although not specified in REST, using HTTP and JSON is the most common practice\,\cite{Neumann2021}.

\parhead{Documentation}
\looseness=-1
A frequently mentioned factor is the clear and comprehensive documentation of REST APIs (\iI{I\textsubscript{1-3}}, \iI{I\textsubscript{5}}, \iI{I\textsubscript{7-8}}, \iI{I\textsubscript{11-12}}), notably mentioned by less experienced interviewees.
Among others, formulated as an obligatory requirement: \imeier{It needs to be documented so that someone can use it}.
The documentation should consist of resources being provided in a clear and convenient way (e.g., not nesting resource paths).
To this end, \zimmer stated that
\iempty{it has always annoyed me when APIs are poorly documented. There are enough of them, then you get a Word document thrown over the fence and then it's not a structured API}.
Ideally, as mentioned by five interviewees (\iI{I\textsubscript{1-3}}, \iI{I\textsubscript{8}}, \iI{I\textsubscript{12}}), documentation uses Swagger/OpenAPI, i.e.,
\igoerke{maybe also OpenAPI, there are also documentation tools like Swagger, (...), provide visual documentation for consumers}.

\parhead{Intuitively usable}
Intuitively usable REST APIs are a major concern of six interviewees (\iI{I\textsubscript{1}}, \iI{I\textsubscript{4}}, \iI{I\textsubscript{7}}, \iI{I\textsubscript{11}}, \iI{I\textsubscript{15-16}}), primarily providers of REST APIs.
While \iI{I\textsubscript{4-5}} and \iI{I\textsubscript{15-16}} phrase it as being straight-forward or easy to use, i.e., \iheppekausen{
	I want it to be easy to use}, further two interviewees (\iI{I\textsubscript{4-5}}) explicitly mention intuition, i.e., \iklaas{that everything is as intuitive as possible}.
Five interviewees (\iI{I\textsubscript{5}}, \iI{I\textsubscript{7}}, \iI{I\textsubscript{11}}, \iI{I\textsubscript{15-16}}) also stated that an API should be self-explanatory and request that \itorkuhl{with common sense I can find my way around}, that
\ikumbarji{I can look at them [the APIs] and understand them directly without having to read through documentation}, or \iheppekausen{that the interface is clear and simple, and understandable (...)}.
The optimal usage of a REST API should be apparent from its design, particularly \imueller{that I can discover them without documentation. (...) 
when I access a resource, I should see links to things that I can do or query}.
Finally, \meier focused on APIs and data formats, emphasizing that they should be \iempty{human-readable, i.e., a REST API with XML or JSON (...).}

\parhead{Consistency of the API}
\looseness=-1
Particularly consultants (\iI{I\textsubscript{2}}, \iI{I\textsubscript{4-5}}, \iI{I\textsubscript{7}}), but also one customer (\iI{I\textsubscript{12}}), emphasized that different APIs and single service API endpoints should be written in a consistent manner.
\rasbach highlights to \iempty{simply get difficulties when using it [an API], because I have the feeling that it is not consistent.}
Particularly, there should be \izang{uniform rules for resource naming} and that if \itorkuhl{I have an endpoint somewhere that I can use, and I get an endpoint [in the response of the latter one] that is used in the same way}.

\parhead{Conciseness of the API}
\looseness=-1
Four interviewees (\iI{I\textsubscript{1}}, \iI{I\textsubscript{6}}, \iI{I\textsubscript{14-15}}) emphasized API conciseness in terms of separation of services, i.e., \imeier{to ensure separation of systems in my REST API}.
To that end, each API endpoint should have a precise and well-focused purpose, because
\isimon{what often makes it [a REST API] more usable now is that teams have been exposed to how to slice and dice them, and the focus has been sharpened to make APIs smaller}.

\parhead{Frameworks}
\looseness=-1
Three interviewees (\iI{I\textsubscript{1}}, \iI{I\textsubscript{3}}, \iI{I\textsubscript{10}}) emphasized the benefits of dedicated tools supporting the design of REST APIs, since \igoerke{REST is very native in that it can be connected to tools.}
To this end, all three interviewees explicitly mentioned OpenAPI and noted that related REST API frameworks come with several features to enhance the usability of APIs, e.g., \igoerke{documentation tools like Swagger (...) provide visual documentation for consumers}.

\parhead{Application Context}
\looseness=-1
\iI{I\textsubscript{4}} and \iI{I\textsubscript{5}} emphasized that REST API usability depends on the usage context of the API.
For example, APIs can be consumer- or developer-oriented, and therefore, it \irasbach{depends a bit on the context how simple or complicated an API is}.

\parhead{API Stability}
\looseness=-1
Finally, \iI{I\textsubscript{6}} and \iI{I\textsubscript{12}} stated that APIs should be stable, expecting \ivoss{stability from the API even if the application behind it may change}---one of the key concepts of REST\,\cite{fielding_dissertation_2000}.
While not explicitly mentioned in the interviews, standards such as semantic versioning\,\cite{PrestonWerner2023} can help to clearly indicate braking changes.

\subsection{Best Practices for Usable REST APIs}
\begin{figure}
	\centering
	\scriptsize
	\begin{tikzpicture}
		\begin{axis}[
			xbar, xmin=0,
			symbolic y coords={Explicitly Considering Security,Testing the APIs,Reviews of the APIs,Proper Documentation,Using OpenAPI, Using Guidelines},
			ytick=data,
			nodes near coords, nodes near coords align={horizontal},
			bar width=0.13cm,
			height=2.9cm, 
			width=6.5cm,
			enlarge y limits=0.14, 
			]
			\addplot coordinates {(10,Using Guidelines)(7,Using OpenAPI)(7,Proper Documentation)(6,Reviews of the APIs)(5,Testing the APIs)(3,Explicitly Considering Security)};
		\end{axis}
	\end{tikzpicture}
	\vspace{-12pt}
	\caption{Best practices for designing usable REST APIs}
	\label{fig:apis:bestpractices}
	\vspace{-16pt}
\end{figure}
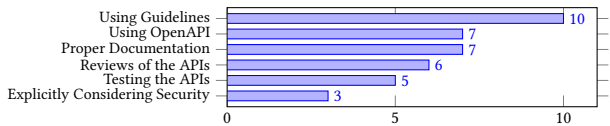

\noindent\looseness=-1
Having identified the factors that influence usability, we asked how to support the creation of usable REST APIs (see \cref{fig:apis:bestpractices}).

\parhead{Guidelines}
With 10 out of 16 interviewees, the majority stated some form of guidelines as a best practice to help developers design usable APIs.
However, only \iI{I\textsubscript{4}}, \iI{I\textsubscript{7-8}}, \iI{I\textsubscript{12}}, \iI{I\textsubscript{14}}, and \iI{I\textsubscript{16}} explicitly named guidelines, while \iI{I\textsubscript{3}}, \iI{I\textsubscript{5}}, and \iI{I\textsubscript{11}}, \iI{I\textsubscript{13}} stated that \igoerke{there are enough standards that I wanted to adhere to}, i.e., recommendations such as \imeyer{how are URLs structured (...)}.

\parhead{OpenAPI}
\looseness=-1
The systematic API specification using OpenAPI is an essential best practice for six consultants (\iI{I\textsubscript{1-4}}, \iI{I\textsubscript{6}}, \iI{I\textsubscript{9}}) and one customer (\iI{I\textsubscript{12}}).
\meier stated that in \iempty{Swagger or in newer OpenAPI 3.0, I find excellent support}.
Among others, allowing them \ivoss{to generate the specification immediately during coding}, which can then be given to consumers.
This is seen as useful because \irunschke{when I have an OpenAPI specification (...) at hand, I can generate my client}.
It also serves as public documentation, so an OpenAPI specification \iper{is something that one should have (...)}.

\looseness=-1 
\parhead{Documentation}
Seven interviewees (\iI{I\textsubscript{5-7}}, \iI{I\textsubscript{11-12}}, \iI{I\textsubscript{14-15}}) consider proper REST API documentation a best practice, but only \iI{I\textsubscript{6}} and \iI{I\textsubscript{12}} named OpenAPI as an essential means to generate documentation.
The rest (\iI{I\textsubscript{5}}, \iI{I\textsubscript{7}}, \iI{I\textsubscript{11}}, \iI{I\textsubscript{14-15}}) states that one should create a \irasbach{good and simple documentation}, not specifically mentioning any technology.
That REST APIs are documented is assumed, i.e., in statements like \itorkuhl{he will probably have provided documentation on this}.

\looseness=-1
\parhead{Reviews}
Six interviewees (\iI{I\textsubscript{2}}, \iI{I\textsubscript{10}}, \iI{I\textsubscript{12-13}}, \iI{I\textsubscript{15-16}}), five of them experienced customers, see reviews as important to improve quality aspects such as usability, i.e., to \imeyer{look over the API design (...) to build uniformity in the company on how REST APIs are built}.
For three of them, adherence to guidelines is an essential aspect of the reviews.
It is essential to give developers feedback, e.g., by \ikumbarji{noting any irregularities [related to guidelines] and then make an appointment with the developers}.

\parhead{Tests}
\looseness=-1
The testing of the REST APIs is described by five interviewees as a necessary step in the development process (\iI{I\textsubscript{1}}, \iI{I\textsubscript{4}}, \iI{I\textsubscript{7}}, \iI{I\textsubscript{12}}, \iI{I\textsubscript{16}}), i.e., \imeier{the entire API has to be tested}.
Thereby, it is essential to have different test levels, among others \iper{unit and integration tests}.

\parhead{Security}
\looseness=-1
Finally, three interviewees (\iI{I\textsubscript{1-2}}, \iI{I\textsubscript{15}}) emphasized the need to explicitly consider API security, stating that \imeier{security must be ensured}.
To this end, a \izang{uniform security in the application landscape} is needed with respect to usability.

\quotebox{RQ1: REST API Usability}{
The main concerns on API usability are adherence to conventions and proper API documentation.
APIs should be designed intuitively usable, consistent, and concise.
Guidelines are seen as essential for designing usable APIs.
Specifically, OpenAPI should be used to specify APIs, also since it allows to generate documentation and client code.
While consultants focus more on such technical solutions, customers seem to be more concerned with processes.
}{}\vspace{-8pt}
\section{Adoption of REST API Guidelines (RQ2)}
\noindent
Institutions have created various REST API guidelines, which (as confirmed above) are an essential factor in implementing usable REST APIs.
Given that REST API usability remains suboptimal, we examined how guidelines are used, what factors influence their successful adoption, and best practices for guideline adoption.

\subsection{REST API Guideline Adoption in Practice}
\noindent
First, we assessed the current state of guideline adoption. 
Of the 16 interviewees, 14 are actively using internal or public guidelines.

\parhead{Provision of Guidelines}
Half of the interviewees (\iI{I\textsubscript{2}}, \iI{I\textsubscript{6}}, \iI{I\textsubscript{8}}, \iI{I\textsubscript{12-16}}) access guidelines through internal wikis, i.e., Confluence, while interviewees (\iI{I\textsubscript{1}}, \iI{I\textsubscript{4-5}}, \iI{I\textsubscript{12}}, \iI{I\textsubscript{16}}) use a website. \iI{I\textsubscript{1}}, \iI{I\textsubscript{3}}, and \iI{I\textsubscript{5}} (also) use PDFs, which according to one interviewee are more like cheat sheets.
\iI{I\textsubscript{3}} notes that interactive media, such as websites, is better than static media such as PDFs.
\iI{I\textsubscript{5}} stated that they use the website of an external guideline and the media provided there.

\parhead{Adherence to Guidelines}
\looseness=-1
Eleven interviewees (\iI{I\textsubscript{1-4}}, \iI{I\textsubscript{6-7}}, \iI{I\textsubscript{11-15}}), four of them mentioned guidelines as best practice, indicated that guidelines are partially avoided or rarely used.
\iI{I\textsubscript{1-2}} and \iI{I\textsubscript{8}} do not utilize guidelines for release, but document their functionality via Swagger/OpenAPI, while one states usage of guidelines as templates when publishing APIs.
One interviewee complained about a lack of automation for guidelines and respective checks.
Another interviewee reports to use guidelines for release documentation and specification, when deploying new APIs.

\looseness=-1
\parhead{Enforcement}
The subliminal wish in most interviews (\iI{I\textsubscript{1-2}}, \iI{I\textsubscript{4-8}}, \iI{I\textsubscript{10-12}}, \iI{I\textsubscript{14-15}}) was that developers should not be forced to follow a guideline, i.e., \iwilde{you should not force it on people, but
you (...) have to teach people}.
\iI{I\textsubscript{8}}, \iI{I\textsubscript{10}} and \iI{I\textsubscript{12}} wish for training or support for questions and problems.
In contrast, \iI{I\textsubscript{12}} stated that ``\textit{the more specific it [a guideline] is, the more likely it is to be violated and if we do not force it, then it is kind of pointless}.''
Finally, one interviewee wishes for an external controlling authority.
While these are more ideas and wishes, they reflect the state reported by four interviewees (\iI{I\textsubscript{2}}, \iI{I\textsubscript{13-15}}) that guidelines are generally poorly enforced.
\iI{I\textsubscript{3-5}} and \iI{I\textsubscript{15}} stated that their guideline cannot be enforced all the time because it contains too many or specific rules.
One interviewee regularly checks APIs and applications for compliance.
Three interviewees (\iI{I\textsubscript{3}}, \iI{I\textsubscript{5}}, \iI{I\textsubscript{8}}) only looked at guidelines from a catalog once, i.e., used it to implement a new API or learn about internal standards. After that, the guidelines were intuitive enough to remembered them.

\subsection{Factors for Successful Guideline Adoption}
\begin{figure}
	\centering
	\scriptsize
	\begin{tikzpicture}
		\begin{axis}[
			xbar, xmin=0,
			symbolic y coords={Tool Support,Maintenance,Objective \& Justification,Processes \& Stakeholders,Comprehensibility,Abstraction \& Focus,Size of the Guideline},
			ytick=data,
			nodes near coords, nodes near coords align={horizontal},
			bar width=0.15cm,
			height=3.2cm, 
			width=7cm,
			enlarge y limits=0.1, 
			]
			\addplot coordinates {(11,Size of the Guideline)(7,Abstraction \& Focus)(6,Comprehensibility)(5,Processes \& Stakeholders)(4,Objective \& Justification)(3,Maintenance)(2,Tool Support)};
		\end{axis}
	\end{tikzpicture}
	\vspace{-12pt}
	\caption{Factors influencing REST API guideline adoption}
	\label{fig:guideline:factors}
	\vspace{-12pt}
\end{figure}
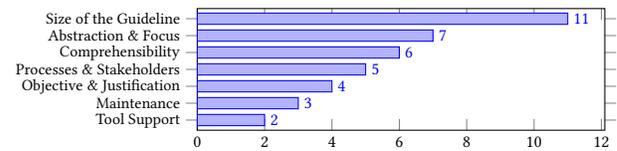

\noindent
While adherence to guidelines improves qualities such as understandability\,\cite{Bogner2023}, as \heppekausen~stated, good guidelines are useless if unused. 
To determine the reasons for the observed opposition to strict guidelines, we identify factors that influence the successful adoption of and adherence to REST API guidelines (see \cref{fig:guideline:factors}).

\parhead{Guideline Size}
For the majority (11 of 16 interviewees (\iI{I\textsubscript{1}}, \iI{I\textsubscript{3-6}}, \iI{I\textsubscript{11-16}})), including all customers, guideline size is the biggest concern in adopting it, stating that
\ivoss{the less text you have the higher the chance of compliance}.
According to them, \imeier{the size of the guideline is the biggest challenge. These are guidelines that are extremely large, and you have to work through them to find out what is relevant for you.},
For example, \irasbach{a known guideline is the one of Zalando (...) They have, (...) 120 to 140 rules, but such a bunch is hard to remember and comply with}.
Almost all interviewees who were aware of popular guidelines (\iI{I\textsubscript{1}}, \iI{I\textsubscript{3-5}}, \iI{I\textsubscript{12-14}}, \iI{I\textsubscript{16}}) noted that they are too extensive, so they use smaller ones in their organization.
When asked about the size of known guidelines, seven interviewees (\iI{I\textsubscript{1}}, \iI{I\textsubscript{3}}, \iI{I\textsubscript{6}}, \iI{I\textsubscript{11-12}}, \iI{I\textsubscript{14-15}}) stated that a guideline needs to be concise, and thus be fast digestible by developers, i.e., one interviewee prefers a two page document.
However, according to four interviewees (\iI{I\textsubscript{1-2}}, \iI{I\textsubscript{6}}, \iI{I\textsubscript{10}}), guideline size is also determined by the technical depth and detail included.
When asked about appropriate size, seven interviewees (\iI{I\textsubscript{1}}, \iI{I\textsubscript{3}}, \iI{I\textsubscript{6}}, \iI{I\textsubscript{10}}, \iI{I\textsubscript{12-13}}, \iI{I\textsubscript{16}}) consider a minimal guideline as adequate in an internal or self-contained environment, as opposed to larger public guidelines.
In practice, there is no fixed size limit above which guidelines are ignored, but it depends on the context of the REST API, i.e., \imeier{I try to evade the big guidelines when writing my APIs, but if I create publicly available ones, I have no other (...)}.

\parhead{Abstraction \& Focus}
Seven interviewees (\iI{I\textsubscript{1-3}}, \iI{I\textsubscript{5-6}}, \iI{I\textsubscript{10}}, \iI{I\textsubscript{16}}) described a good guideline as focused on implementable rules, while too broad guidelines will not be read.
Therefore, \zang prefers selections of rules, because \iempty{you have to see what situation you want to solve and if you need a rule for that.}
Another interviewee states that \irasbach{you should concentrate on important rules and may continue to extend with lesser prioritized rules}.
One interviewee states that this is done through individual guidelines for each department, while another wants just one guideline.
According to \iI{I\textsubscript{1}} and \iI{I\textsubscript{5}}, bad guidelines combine rules that are adverse or unrelated to their context.
Particularly, \meier requests that \iempty{my guideline should help me describe my interface and not contain too many details, which are irrelevant here.}
Related to this, \iI{I\textsubscript{5}} and \iI{I\textsubscript{16}} state, too many technical details or performance enhancements bog down a guideline, since \ikumbarji{the guideline may describe with an example of how to return data, but if I encounter more complicated situations, this example may not be applicable}.
Finally, one interviewee remarked that the level of detail or stage of development determines  a guideline's acceptance.




\looseness=-1
\parhead{Comprehensibility}
Six interviewees (\iI{I\textsubscript{1}}, \iI{I\textsubscript{5-6}}, \iI{I\textsubscript{11-12}}, \iI{I\textsubscript{15}}) state, \imueller{guidelines should be easily comprehensible}.
To this end, a guideline \imueller{should be concise (...) you don't have much time to read through}.
Guidelines should include keywords for every rule, \imueller{at best in headlines}. 
However, according to \iI{I\textsubscript{4}}, guidelines that use buzzwords for explanations without proper sentences, or clear prioritization of rules (another interviewee), are less usable.
Six consultants (\iI{I\textsubscript{1-2}}, \iI{I\textsubscript{4}}, \iI{I\textsubscript{5-6}}, \iI{I\textsubscript{9}}) emphasize to support rules by explanations, examples  and providing technical depth to be informative (\iI{I\textsubscript{1}},\,\iI{I\textsubscript{6}}).

\parhead{Processes \& Stakeholders}
\looseness=-1
For five of six customers (\iI{I\textsubscript{10}}, \iI{I\textsubscript{13-16}}), the processes within an institution and involved stakeholders are an essential factor for the successful adoption.
\heppekausen emphasized the challenge of integrating guidelines into workflows, and another emphasized the challenge of choosing an appropriate guideline.
Particularly concerning stakeholders, \iheppekausen{acceptance is most important. I can write wonderful things in my guideline, but if developers don't accept it, it's useless. I have to connect to the people, consider their different skill levels, and create consensus of the problem}.
According to four interviewees (\iI{I\textsubscript{2}}, \iI{I\textsubscript{13-14}}, \iI{I\textsubscript{16}}), one must consider that \ikumbarji{everyone has their own level of knowledge, you can definitely see, more with older developers, that they develop APIs the same way they did before with SOAP}.
Particularly, \isimon{there is always a problem of old thought patterns---after working for 30 years, knowing your data models and functions, one may not see the problem, that others don't understand them in the same way}.
It is essential to maintain a community with appropriate culture, since it is \imeyer{problem of willingness to learn. Developers who are interested are taking part in meet-ups and do look up rules in forums while others do not.}
\iI{I\textsubscript{10}} and \iI{I\textsubscript{13}} emphasize the need to enforce guidelines in some way, especially to senior developers who tend to be reserved.
Focusing on processes, \wilde states \iempty{If you want to create APIs according to guidelines, you need to make sure, that all developers are able to}.
One interview emphasized the appropriate choice for the guideline medium as factor.


\parhead{Objective \& Justification}
\looseness=-1
According to \iI{I\textsubscript{1}}, \iI{I\textsubscript{3}}, \iI{I\textsubscript{5}}, and \iI{I\textsubscript{10}}, guidelines must serve a clear purpose 
and \iwilde{explain why rules exist}.
To this end, plans and explanations for the aimed at maturity level (cf. Richardson Maturity Model
\,\cite{maturitymodel}), i.e., which concepts of REST are targeted to be implemented, should be included. 
A guideline should not be self-purpose. 
It must enhance the whole workflow, in which enforcement is handled (\iI{I\textsubscript{6}}, \iI{I\textsubscript{10}}).

\parhead{Maintenance}
\looseness=-1
For three interviewees (\iI{I\textsubscript{1-2}}, \iI{I\textsubscript{7}}) it is essential that guidelines are up-to-date.
One interviewee noted that regardless of the scope, guidelines must be kept up to date, and one interviewee emphasized that an outdated guideline is bad.
Overall, the maintenance of old guidelines is an important factor that affects their practical adoption for three interviewees (\iI{I\textsubscript{1}}, \iI{I\textsubscript{7}}, \iI{I\textsubscript{15}}).
In addition, 
one interviewee emphasizes that any examples given in a guideline should work. 
One interviewee suggests a REST API guideline should promote feedback and revisions from users and developers. 

\parhead{Tool Support}
\looseness=-1
Two consultants remarked that a usable guideline provides tool support, or is at least machine-readable, i.e., OpenAPI (\iI{I\textsubscript{1}}, \iI{I\textsubscript{5}}).
The size of a guideline matters less, if a sufficient support is given by the guideline  creators like Zalando (one interviewee), or if machines and tools can provide feedback to the developers (one interviewee).
Lastly it was remarked by one interviewee that a guideline should also have human support.

\subsection{Best Practices for Adopting Guidelines}
\noindent
We identified six best practices for adopting guidelines.

\looseness=-1
\parhead{Community Work}
When adopting guidelines, it is essential to convince all stakeholders of its benefit.
It is essential \iper{to proactively promote guidelines} and \isimon{to support teams that have not worked with it}.
The ultimate goal is to build \iheppekausen{a community on this topic}.
Instead of management strictly enforcing a guideline, it is essential \imeyer{to convince why this way [guideline rule] is the standard or simpler way}.
Courses, e.g., on API design, can be a suitable means.
Following the broken window theory of technical debt\,\cite{leven_broken_window}, developers who notice that their coworker's API ignores the guideline may also stop complying.

\parhead{Tools and Automation}
Although continuous integration has automated many software development tasks, compliance with API guidelines is still primarily checked manually, i.e., by integration teams (\kumbarji) or during certifications and code reviews (\per).
However, interviewees see automated checking for compliance with REST API guidelines as highly desirable, i.e., \iwilde{if you not only invest in your guideline but also automation, then developers wouldn't need to read and remember all rules}.
\iI{I\textsubscript{11}} reports positive experiences with automation---\imueller{we had hundreds of microservices, and you just cannot have a process to check the APIs. That is why we developed an API linter to support the use of guidelines}.
This is in line with current trends, since \iwilde{typically today, teams are working on automating processes, mostly by automating reviews over time. Step by step  more rules get checked against guidelines}.

\looseness=-1
Examining the detection rules of static analyzers such as SonarQube\,\cite{sonar,sonar-openapi} reveals that REST API quality rules mostly assume the use of tools such as OpenAPI.
Besides using OpenAPI being considered a best practice (\iI{I\textsubscript{1}}, \iI{I\textsubscript{5}}), this further encourages its use.
However, this reveals also a gap in the current landscape of API linters: While tools excel at syntactic and schema validation\,\cite{Sundberg2025}, they struggle with usability and context-dependent guidelines that require human judgment\,\cite{Rauf2019}.
Usability testing remains a manual or semi-automated process, essential for ensuring APIs are intuitive and user-friendly.
In this context, REST API guidelines can be categorized into general principles (e.g., statelessness, caching), which are difficult to automate, and fine-grained rules (e.g., naming, schema validation), which are more amenable to automated checking\,\cite{Coblenz2023}.

While tool support can reduce cognitive load, it may also reduce commitment to guidelines, raising questions such as \iklaas{is there already something implemented in the build pipeline?}.
To address this, organizations should complement automation with educational scaffolding\,\cite{Quintana2002}, such as interactive feedback, that reinforces the purpose of guidelines.
For example, linters could not only flag violations but also explain the impact of non-compliance (e.g., ``This naming convention improves discoverability for client developers''). 


\parhead{Interactive Guidelines}
A key recommendation for guideline provision is to utilize interactive media.
Ten interviewees (\iI{I\textsubscript{1-2}}, \iI{I\textsubscript{6}}, \iI{I\textsubscript{8}}, \iI{I\textsubscript{11-16}}) effectively access guidelines through internal wikis, such as Confluence, i.e., \izang{our own guideline for our REST APIs, documented in a wiki format}.
Interactive media is seen as superior to static documents, providing an engaging, user-friendly experience.
In particular, interactive navigation \imeier{where I can jump back and forth and search better} is important.
It enhances accessibility and usability, facilitating better compliance with guidelines.
Thereby, priority keywords for every rule in the guideline can provide a quick fist overview, since \imueller{in everyday use, you don't have much time to read through}.
Using priority keywords, as in the Zalando guideline\,\cite{zalando}, helps users understand and adhere to guidelines.

\parhead{Appropriate Rule Selection}
\looseness=-1
All interviewees agreed on have an appropriate guideline for the organization is most important.
To ensure optimal results, it is recommended to adopt a concise guideline.
The interviewees stressed that the size of a guideline is the biggest concern.
Recall that \voss noted, that \iempty{the less text you have, the higher the chance of compliance}.
Consequently, organizations should use smaller guidelines internally to enhance adherence.
A best practice for ensuring adherence to guidelines is to design them to be intuitive and focus on organization relevant rules.

\quotebox{RQ2: Adoption of REST Guidelines}{
\hspace{.5pt}
The biggest challenge in adopting guidelines is the acceptance by developers, since, as \rasbach summarized, guidelines are extra work for developers.
The primary guideline-specific factors impacting their adoption are their size and closely related to this their provided abstraction and focus.
The best way to ensure adoption is to convince developers of the benefits.
This suggests that institutional culture and processes are key factors in improving adoption.
Tooling can particularly help embed guidelines into development processes and automate compliance checking to some extent.}

\section{Creating Custom Guidelines (RQ3)}
\noindent
The factors for successful adoption of REST API guidelines indicate multiple organization-specific factors and that public guidelines are often too broad.
Thus, we inspected the motivations, challenges, and best practices behind custom guidelines.

\subsection{Motivations for Custom Guidelines}
\noindent\looseness=-1
To understand the motivations for custom guidelines in the presence of public guidelines, we further asked for reasons.

\parhead{Minimize Size}
As the above discussion shows, guideline size is a major factor in successful adoption, and public guidelines are seen as too big.
Therefore, a main motivation for custom guidelines is to minimize size to increase efficiency and effectiveness.
Still, public guidelines are often considered, but they want to \imeier{delete what is not relevant} to \iper{keep it simple}.
Consequently, organizations often create \imeier{a slimmed down version} of guidelines.
The factors below play an essential role in determining which rules to include.

\parhead{Domain-specific Context}
\looseness=-1
A custom guideline is described as necessary by five interviewees (\iI{I\textsubscript{1}}, \iI{I\textsubscript{6-7}}, \iI{I\textsubscript{10}}, \iI{I\textsubscript{13}}), because it must fit company and industry context.
One has to consider that each \imeier{interface is unique. Therefore, my guideline should reflect all relevant aspects}.
In particular, rules of public guidelines do not apply to every context, so organizations want to remove them since \ivoss{irrelevant rules reduce acceptance}.
Context dependent, organizations often \imeyer{notice that there are some aspects that one does not have to treat (...)}
in their application context, allowing to reduce the size while increasing the chances of compliance.
Additionally, \iI{I\textsubscript{1}}, \iI{I\textsubscript{3}} and \iI{I\textsubscript{4}} mention the concrete technologies used as context, especially when changed, the guideline needs to be tailored to stay relevant. 

\parhead{Organizational Priorities}
\looseness=-1
Seven interviewees (\iI{I\textsubscript{1}}, \iI{I\textsubscript{4}}, \iI{I\textsubscript{6}}, \iI{I\textsubscript{11-12}}, \iI{I\textsubscript{14-15}}) noted that organizational priorities shape API guidelines.
In the end, \iper{this is a matter of taste, with our [the organization] own flavor}.
Organizations want to reflect custom API styles that stem from their custom priorities and unique selling points.
Particularly, a guideline should reflect the organizations priorities, i.e., \iheppekausen{we then narrowed it down [the Zalando guideline] to 10 points and described them using [organizational] wording and limited ourselves to it}.

\parhead{Organizational Processes}
Another essential motivation for customization is that a guideline has to \iwilde{fit processes} of the adopting organization.
One interviewee states that the Zalando guideline is not flexible enough and another that guidelines need an appropriate \iper{balance between governance and team autonomy}.
How a guideline reflects this is specific to each organization and therefore requires customization.
To increase autonomy, organizations customize guidelines by deciding what to regulate, i.e., they \itorkuhl{agreed on that [guideline rules] and left the rest to the particular situation}.

\parhead{Reflect Discussions}
\looseness=-1
An essential factor for a successful guideline is the \iwilde{mutual agreement on what is good}.
Often there are \iwilde{different possible solutions}, but only \iwilde{one must be selected}.
Selection is often the result of internal discussions that \ivoss{involve developers}.
E.g., \ikumbarji{when you face a problem or question (...) we search for a solution in the team}.
The guidelines should document the outcome to support future decisions, therefore guidelines should \izang{reflect discussions}, automatically leading to customization.
Furthermore, there is often \imueller{feedback of development teams} that should be addressed in the guideline, i.e., on what to describe better or how well rules can be realized in practice.
Particularly, customization allows to \isimon{address that rules are not followed}.

\parhead{Control over Changes}
\looseness=-1
Finally, \iI{I\textsubscript{2}}, \iI{I\textsubscript{12}}, and \iI{I\textsubscript{16}} indicate that having control over changes can be a motivation. 
Firstly, organizations want to keep their guidelines up to date, because \ikumbarji{you cannot just develop according to the same guidelines for ten years. That is why you always have to see what the current state of science and practice is}.
Still, organizations do not want to rely on external guidelines being maintained.
Secondly, they need stable guidelines and want to limit changes, where guideline size is a factor to consider, \iper{because it is manageable and stable due to the few rules}.

\subsection{Guideline Creation in Practice}
\noindent\looseness=-1
Of the 16 interviewees, seven helped create a single guideline, one contributed to several, six had minimal involvement, and two were not involved at all.
Thus, we asked these 14 interviewees how guidelines are created in their organizations.

\parhead{Reference Guidelines}
\looseness=-1
When asked why they create a custom guideline instead of using an existing public guideline, \iI{I\textsubscript{1}}, \iI{I\textsubscript{4}}, and \iI{I\textsubscript{10}} agreed that \iklaas{people already thought about it and in the best case, I don't have to figure it out myself}.
Therefore, the creation itself is based on other guidelines utilized as templates (\iI{I\textsubscript{1}}, \iI{I\textsubscript{10}}, \iI{I\textsubscript{15}}), acquired through web research or other companies.
Four interviewees raised criticism in the way how custom guidelines are created (\iI{I\textsubscript{1}}, \iI{I\textsubscript{5}}, \iI{I\textsubscript{7}}, \iI{I\textsubscript{11}}), i.e., that \irasbach{it is a bad process to create your guideline by randomly copying and crafting together a wild bunch of rules, since they can contradict one another}.
Existing guidelines should not just be copied, but definitely help by giving inspiration.
Some rules may be copied if they seem helpful, but rules can conflict with other rules.
To the point of inspiration, one statement was that \iwilde{for some time the Zalando guideline was (...) a baseline for tailoring an individual guideline, but lastly I heard that this trend shifted to the Google guideline}.

\parhead{Contributors}
\looseness=-1
While four interviewees stated that software architects, quality assurance, and developers of the API (\iI{I\textsubscript{1-2}}, \iI{I\textsubscript{6}}, \iI{I\textsubscript{11}}) are involved in creating guidelines, three interviewees (\iI{I\textsubscript{1}}, \iI{I\textsubscript{12}}, \iI{I\textsubscript{14}}) stated to ensure that everyone in the project is integrated.
Other stakeholders involved in guideline creation include technical leads as stated by \iI{I\textsubscript{1}} and \iI{I\textsubscript{11}}.
In addition, single interviewees indicated to involve also API experts, security experts, API consumers, the distribution team, and team leads.
One interviewee remarked that any guideline creation or modification is done by the core team itself.
\iI{I\textsubscript{5}} and \iI{I\textsubscript{14}} mentioned that any changes made to the guideline need to be approved by the developers before changes are made.

\parhead{Data Collection Process}
\looseness=-1
Regarding the process of creating guidelines, \iI{I\textsubscript{2}}, \iI{I\textsubscript{6}}, \iI{I\textsubscript{10}}, and \iI{I\textsubscript{12}} refer to their agile work environment, which ensures a feedback loop with the participants. 
Interviewees \iI{I\textsubscript{2}}, \iI{I\textsubscript{11}} and \iI{I\textsubscript{15}} mention that they first acquire knowledge through research, and after gathering some guidelines, modify them according to their needs.
\per reported they have an explicit \iempty{RFC process [Request For Comments], so if someone has a request, you can put it in the wiki}.

\parhead{Communication}
When guidelines change, 9 out of 15 interviewees actively notify users of their guideline or API (\iI{I\textsubscript{1}}, \iI{I\textsubscript{5-7}}, \iI{I\textsubscript{12-16}}).
To this end, they use Blog posts, emails, or internal communication channels, e.g. \ivoss{we have an announcement mechanism that uploads to our website (...) We also tried Teams channels (...) and serial emails}.
\iI{I\textsubscript{1}} and \iI{I\textsubscript{11}} reported that emails work well.
\iI{I\textsubscript{1}} uses a personal channel to \imeier{inform them [the guideline consumers] actively through email, about changes and when they apply} and \iI{I\textsubscript{11}} uses \imueller{newsletters to announce changes}, particularly stating that this works well.
One conclusion was to \imueller{always over-communicate over several channels}.
Lastly, two interviewees (\iI{I\textsubscript{7}}, \iI{I\textsubscript{11}}) communicate any alterations via tool support, either through REST endpoints or linters.
In contrast, the guideline version is silently updated by four participants (\iI{I\textsubscript{1}}, \iI{I\textsubscript{13}}, \iI{I\textsubscript{15-16}}). \iI{I\textsubscript{4}} and \iI{I\textsubscript{16}} do not communicate changes.

\parhead{Continuous Maintenance}
\looseness=-1
Three interviewees (\iI{I\textsubscript{1}}, \iI{I\textsubscript{5-6}}) conclude that a guideline is never finished, e.g., due to the need for addition and revision of new rules.
In this regard, two \iI{I\textsubscript{1}} and \iI{I\textsubscript{5}} emphasized that guideline development needs to be actively terminated, either by stopping or completing the corresponding API development.
\iI{I\textsubscript{6}} stressed the importance of maintaining a guideline throughout its life-cycle.
Particularly, guidelines reflect the decisions in the project, and therefore, organizations usually \ivoss{start with a small set [of rules] that everyone can quickly agree on and then see how well they work and gradually expand them. That's why they are never finished.}

\subsection{Challenges}
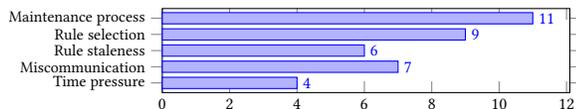
\begin{figure}
	\centering
	\scriptsize
	\begin{tikzpicture}
		\begin{axis}[
			xbar, xmin=0,
			symbolic y coords={Time pressure,Miscommunication,Rule staleness,Rule selection,Maintenance process},
			ytick=data,
			nodes near coords, nodes near coords align={horizontal},
			bar width=0.15cm,
			height=2.7cm, 
			width=7cm,
			enlarge y limits=0.15, 
			]
			\addplot coordinates {(11,Maintenance process)(9,Rule selection)(6,Rule staleness)(7,Miscommunication)(4,Time pressure)};
		\end{axis}
	\end{tikzpicture}
	\vspace{-14pt}
	\caption{Challenges to the creation of custom guidelines}
	\label{fig:guideline:creationchallenges}
	\vspace{-12pt}
\end{figure}

\noindent\looseness=-1
To better understand why current practices of tailoring guidelines to the organization to increase adoption are not resulting in optimal compliance, we asked the interviewees about the challenges of creating custom REST API guidelines (see \cref{fig:guideline:creationchallenges}).

\parhead{Maintenance process}
\looseness=-1
The maintenance process is seen as a challenge by most interviewees (11 of 14 interviewees that create guidelines: \iI{I\textsubscript{1-2}}, \iI{I\textsubscript{5-6}}, \iI{I\textsubscript{10}}, \iI{I\textsubscript{11-13}}, \iI{I\textsubscript{15-16}}).
Since a guideline cannot include everything, two interviewees (\iI{I\textsubscript{7}}, \iI{I\textsubscript{11}}) mention that surprising or new situations need be included after occurence.
\iI{I\textsubscript{10-11}} pointed out that the alteration process itself poses challenges, specifically regarding the consensus and implementation.
Backwards compatibility has to be ensured, and no breaking changes should be made. 

\parhead{Rule selection}
\looseness=-1
Rule selection is seen as a challenge by nine interviewees (\iI{I\textsubscript{1}}, \iI{I\textsubscript{5-7}}, \iI{I\textsubscript{10-11}}, \iI{I\textsubscript{14-16}}).
Three interviewees (\iI{I\textsubscript{1}}, \iI{I\textsubscript{7}}, \iI{I\textsubscript{10}}) stated that it is important but challenging to keep and provide the context of the API, particularly in guidelines with manageable size.
Recall, there is considerable resistance to reading through long texts; however, it is a challenge to select a set of appropriate rules to include in the guideline.
In relation to this, two interviewees mentioned, that preventing inconsistencies 
is challenging (\iI{I\textsubscript{5}}, \iI{I\textsubscript{11}}).
Also, deciding on an appropriate level of abstraction and including good and appropriate examples is challenging.
This is further challenged by the circumstances that \irasbach{there are many points where there is simply no clear best solution (...)}. 
This usually leads to discussions of which solution to adopt in an organization., i.e., adding it to the guideline.

\looseness=-1
\parhead{Rule Staleness}
Six interviewees (\iI{I\textsubscript{1-2}}, \iI{I\textsubscript{7}}, \iI{I\textsubscript{12}}, \iI{I\textsubscript{15-16}}) see keeping rules up to date as a challenge.
Guideline initiatives can fail from one stale rule alone, considering that \iheppekausen{if a developer encounters outdated guideline rules, he or she will not read through the rest, thinking those rules will be outdated too}.
However, guidelines do not evolve as fast as the technology stacks involved, i.e., \per stated that he has not \iempty{seen substantial changes to the guideline in the last three years. But there are other guidelines where it is different, especially, testing strategies where discussions and changes occur more frequently}.

\parhead{Miscommunication}
While communication was seen as essential, seven interviewees identified good communication of the guideline and its changes as a significant challenge (\iI{I\textsubscript{1}}, \iI{I\textsubscript{5-6}}, \iI{I\textsubscript{10-11}}, \iI{I\textsubscript{15-16}}).
Organizations must avoid situations in which \ivoss{someone uses the old guideline}, but providing a sensible versioning scheme is major challenge for three participants.
The involved communication is dependent on the organization, especially size, and what software or method reaches most people.
While providing active notice of new guidelines or changes in a guideline, two interviewees (\iI{I\textsubscript{6}}, \iI{I\textsubscript{11}}) note that used channels tend to be spammed and notifications are lost. For example, \voss reports that they \iempty{have an announcement mechanism that uploads to our website, but we want to change that. There is always someone who does not see it. We also tried channels or emails with the same result and it developed a spam characteristic}.

\parhead{Time pressure}
According to \iI{I\textsubscript{5-7}} and \iI{I\textsubscript{11}} one factor is the time-consuming nature of creating guidelines. 
When asked about workload or costs, \mueller answered that \iempty{especially in the start-up field, you need to think about time-to-market. (...) If there is no time to create guidelines, you just copy other guidelines like Zalando.}
This challenge was also phrased as an investment decision, stating that \itorkuhl{creating a guideline is unpleasant work (...). 
Agreeing on the same language costs time initially, but you become faster in the future}. 

\subsection{Best Practices}
\noindent
From the answers presented and further in-depth questions, we derived best practices for custom guidelines that improve REST APIs and facilitate compliance while keeping costs low.

\parhead{Start Small}
\looseness=-1
Given the size of many guidelines and the difficulty of grasping and remembering all rules, starting with a smaller or focused subset is beneficial. 
It is important because \irasbach{those responsible have to read and internalize the guidelines first.
Depending on the length of the document, this is a major criterion}.
In the end, the interviewees' experience that \iheppekausen{if you start with a small guideline, you will have good acceptance because the developers are directly involved and if there are specific questions, you can go into more detail}.
This allows to reflect situations and challenges observed during REST API development and keeps rules relevant.
Since the interviewees emphasized the consistency of REST APIs as an important quality aspect, although not explicitly stated by them, this has to be considered in the rule selection.
While improving only changed code is best practice\,\cite{Digkas2022}, ensuring API consistency may require refactoring the entire REST API when adding a new rule.

\looseness=-1
\parhead{Collective Ownership}
Involving users in the guideline creation prior to its widespread adoption is seen as an best practice to improve guideline acceptance (\zang, \voss, \per, \meyer, and \simon).
A key benefit is that \isimon{with involvement we get the advantage to say, you did accept this, if you do not comply we need to change it. So we are not seen as the team that dictates rules.}.
It also allows to deal with the challenge that \ikumbarji{not everyone agrees on guideline rules, just because we implemented them. We are not total experts and developers have their own knowledge on the topic}.
Therefore, organizations should involve developers to leverage existing knowledge and as they have to commit to the guideline afterward.
As a positive example, \voss particularly pointed out that \iempty{in the past, we organized the guidelines centrally. Now, we try building it with a community approach and integrate people earlier.}
While the approach to integration differs due to the development teams' sizes, the basic idea is to give guideline users the opportunity to participate.
Changes to the guideline can be suggested or required by different sources, i.e., new situations can be encountered or developers suggest changes based on newly gathered experience.
When asked why and when guideline rules have to be changed, \simon stated that \iempty{when rules are not complied with, there is reason to talk about. The legal framework can change, and we need to see how it affects us. Or simply someone mentions the need for change.}
In line with the interviewees' perception, collective ownership has effective in improving code quality\,\cite{Greiler2015}.
However, different factors impact effectiveness of collective software artifact ownership\,\cite{Koana2024}, which may have to be considered also for REST API guidelines.

\quotebox{RQ3: Creating and Tailoring REST API Guidelines}{
	\hspace{-.4pt}
	Guidelines are typically customized by organizations to 
	tailor them to their needs, trying to increase developer compliance. 
	The main challenge is to actively maintain the guideline to keep it relevant.
	To scope a guideline it is essential to identify the rules that should be added to the guideline---at an appropriate level of abstraction for the targeted stakeholders, while having limited resources. 
	We identified the iterative growth of guidelines combined with the active involvement of all stakeholders 
	as best practices to keep guidelines focused and embedded in the developers minds.
}{}

\section{Threats to Validity}
\noindent

\parhead{Internal Validity}
\looseness=-1
The validity of the interviews could be threatened by factors, such as subject expectancy bias, observer bias, and selection bias.
Especially, subject expectancy bias could have affected the interviews, since the interviewer was known among some interviewees.
The interviewer itself had significant experience in working with REST APIs and guidelines, among others from working for the organization from which the interviewees were recruited.
While this provides him with necessary background, this could affect interviews due to an internal bias.
To mitigate this threat, we involved further authors that are not related to the organization in the analysis of the interview transcripts.
It is important to note that both the questions and answers were translated, since the interviews were conducted in German. This may affect the internal validity due to potential bias in the exact translation of terms.
However, as the full context was always considered, the potential impact is minimal.

\parhead{External Validity}
A threat to the external validity is the limited number of interviewees due to focusing on a smaller group of experts instead of a broader group of general developers.
Nevertheless, saturation was reached in later interviews, confirming all key challenges and practices. 
Another potential threat is the large number of participants who work for the same company.
 Yet, these consultants bring diverse experiences from various clients and do not share REST APIs or guidelines, minimizing bias.
As shown above, the individual experience of developers can influence how they view guidelines.
Although saturation was clearly evident in the data analysis, the details of challenges may be different in other organizations.
Cultural and contextual parameters are likely to play a role.
However, the general challenges and best practices identified are unlikely to change.

\section{Conclusion}
\noindent
REST APIs are core business assets, yet their usability often impedes their usage.
While API guidelines aim to improve usability, their practical application and effectiveness remain understudied. 
To fill this gap, our study examines the factors influencing successful adoption and organizational tailoring of REST API guidelines.

While we found that API usability depends on consistent conventions, intuitive design, and thorough documentation, we also learned that REST API guideline adoption faces developer resistance, as extra rules are often perceived as burdensome.
Successful adoption thus requires demonstrating clear benefits, tailoring guideline scope and abstraction, and embedding compliance into workflows through supportive culture and tooling. 
We found that REST API guidelines are typically tailored to specific organizations to reduce complexity and improve relevance, but maintaining their usefulness demands active upkeep, stakeholder collaboration, and iterative refinement to balance practicality and resource limits---often complicated by conflicts between valid but competing practices. 
Shared ownership and incremental growth emerged as best practices to keep guidelines focused, practical, and integrated into developers’ daily work.
Linting, in particular, offers opportunities not only for enforcing compliance but also for educating developers.



\looseness=-1
Our results give rise to the following research directions.
First, advancing automated guideline compliance checks is crucial, as it appears to be the only scalable strategy for governing numerous, evolving rules across a growing number of API endpoints in a team.
Additionally, while we considered REST API guidelines holistically, analyzing individual rules and their taxonomy could benefit both research and practice to address the problem of limited cognitive capacity.
Finally, recall that guidelines are often very specific to organizations. While work that analyze REST API guidelines and categorize their rules already exists\,\cite{murphy_preliminary_2017}, further work is needed for a unified catalog of universally valid rules.


\bibliographystyle{acm}
\balance
\bibliography{doc_bib}

\end{document}